\documentclass{iauc}
\usepackage{graphicx}

\title[Giant Halos in Dwarf Irregular Galaxies] 
{Giant Halos in Dwarf Irregular Galaxies vs. Dwarf Elliptical Galaxies}
\author[Lee \& Hwang]   
{Myung Gyoon Lee$^1$ and %
Narae Hwang$^1$} 

\affiliation{$^1$Astronomy Program, SEES, Seoul National University, Seoul, 151-742, Korea \\
email: mglee@astrog.snu.ac.kr, nhwang@astro.snu.ac.kr \\[\affilskip]
}

\pubyear{2005}
\volume{198} 
\pagerange{???--???}
\date{?? and in revised form ??}
\jname{Near Field Cosmology with Dwarf Elliptical Galaxies}
\editors{H. Jerjen and B. Bingelli, eds.}
\begin{document}

\maketitle

\begin{abstract}
We present a discovery of a giant stellar halo in NGC 6822, a dwarf irregular galaxy in the Local Group. This halo is mostly made of old red giants, showing striking features:
1) it is several times larger than the main body of the galaxy seen in the optical images,
and 2) it is elongated in the direction almost perpendicular to the HI disk of NGC 6822. 
The structure of this stellar halo looks similar to the shape of dwarf elliptical galaxies,
indicating that the halos of dwarf irregular galaxies share the same origin with those of the dwarf elliptical galaxies.


\end{abstract}
\firstsection 
\section{Introduction}

It has been known long that old stellar populations exist in the dwarf irregular galaxies and dwarf elliptical galaxies and the $I$-band magnitude of the tip of these old populations (red giant branch) has been
often used to estimate their distances (Lee et al. 1993). 
However, little is known about the size and morphological structure of the halos of these old stellar
populations in the dwarf galaxies.
A study of the morphological structure of the halos in dwarf galaxies may provide some clues
to understanding the origin of the dwarf irregular galaxies and the dwarf elliptical galaxies.
We have carried out a project to investigate the halos
in dwarf irregular galaxies using the MegaPrime camera at the CFHT. 
Here we present a case study of NGC 6822 which is a famous dwarf irregular galaxy in the Local Group.
NGC 6822 is known to have a huge rotation HI disk (de Blok \& Walter 2000) and a wide distribution of
carbon stars (Letarte et al. 2002). 
It is an well isolated galaxy belonging neither to our Galaxy nor to M31.
NGC 6822 is located at the distance of 500 kpc, and one arcmin at this distance corresponds to 145.4 pc.

\section{Color-Magnitude Diagram of the Dwarf Irregular Galaxy NGC 6822}

We display in Figure~1 the $i-(g-i)$ color-magnitude diagram of three regions at $R<5'$, $5'<R<10'$, and $10'<R<15'$ where $R$ is the angular distance from the center of NGC 6822.  This is part of the data covering a 1 deg $\times$ 1 deg field including NGC 6822 (corresponding to a field of 8.7 kpc $\times$ 8.7 kpc).
Figure~1 shows several distinguishable stellar populations in NGC 6822 as marked by the boxes: 
blue main sequence stars and blue supergiants (MS), red supergiants (RSG), old red giant branch stars (RGB), asymptotic giant stars (AGB), and young red giants (RG). A broad vertical structure seen at $(g-i)\approx0.9$ represent foreground stars. 

Young stars are mostly seen at $R<10'$, while the old RGB stars are abundant even at $R>10'$.
This old population may belong to the halo of NGC 6822. 
We have used the stars inside each box for the following analysis to investigate the morphological
structure of each population.

\begin{figure}
\includegraphics[scale=0.50]{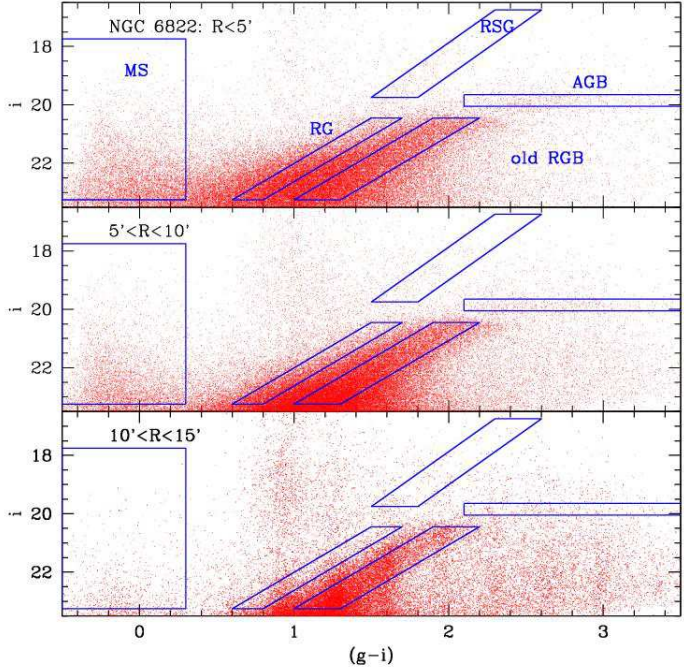}
  \caption{Color-magnitude diagrams of the stars at $R<5'$ (top panel),
$5'<R<10'$ (middle panel), and $10'<R<15'$ (bottom panel) where $R$ is the angular distance 
from the center of NGC 6822.
Thick solid line boxes represent the regions selected for the analysis of the different stellar populations.} 
\label{fig1}
\end{figure}

\section{Spatial Distribution of Stellar Populations in NGC 6822}

We have derived the surface number density maps for the MS, young RG and old RGB in the field of 1 deg $\times$ 1 deg to investigate their spatial distribution, displaying
them in Figure~2  in comparison with the greyscale map of the digitized POSS map of NGC 6822.
Figure~2 shows several striking features.

First, the structure of the old RGB (halo) is dramatically different from the main body ($10' \times 20'$) 
of NGC 6822 seen in the POSS map. It is surprisingly much larger than the POSS image, exceeding one squaredegree field. Its inner region is elongated horizontally, while its outer region is elongated along the NE-SW direction.
Its inner region looks elliptical, very similar to the structure of dwarf elliptical galaxies.
However, its outer part shows some distorted structure in the NE-SW end
indicating some tidal interaction. The existence of this tidal feature is intriguing, 
because NGC 6822 has no known companion galaxies in the nearby distance. 
 
Second, the spatial distribution of the young MS in the central region 
is similar to the shape of the main body of NGC 6822 seen in the POSS map, 
but it is much more extended and elongated along the NW-SE direction. 
This structure is consistent with the HI structure (de Blok \& Walter 2000). 

Third, the spatial distribution of the young RG looks round and similar to that of the central part of the old RGB
distribution, but its center is offset from the center of the RGB distribution. 
In addition, it shows stronger central concentration than that of the old RGB. 
 
We have also investigated the radial variation of the surface number density for the RGB, RG, MS, AGB and carbon stars (Letarte et al. 2002) 
as plotted in Figure~3.
Figure~3 shows that the young populations (MS and RG) are more centrally concentrated than the older
populations (RGB, AGB and  carbon stars).

\section{Conclusion}

Through the deep wide-field imaging we have found that NGC 6822 has a giant halo made of old red giants, extending much farther than the optical body and HI disk of this galaxy. 
The structure of this stellar halo looks similar to those of dwarf elliptical galaxies so that we conclude
that the halos of dwarf irregular galaxies may share the same origin with those of the dwarf elliptical galaxies.
NGC 6822 also is known to have an old globular cluster at as far as 13 kpc from the center, supporting
that its halo is really huge (Hwang et al. 2005). 
The existence of the giant halos may be common among dwarf irregular galaxies, waiting to
be discovered.


\begin{figure}
\includegraphics[scale=0.6]{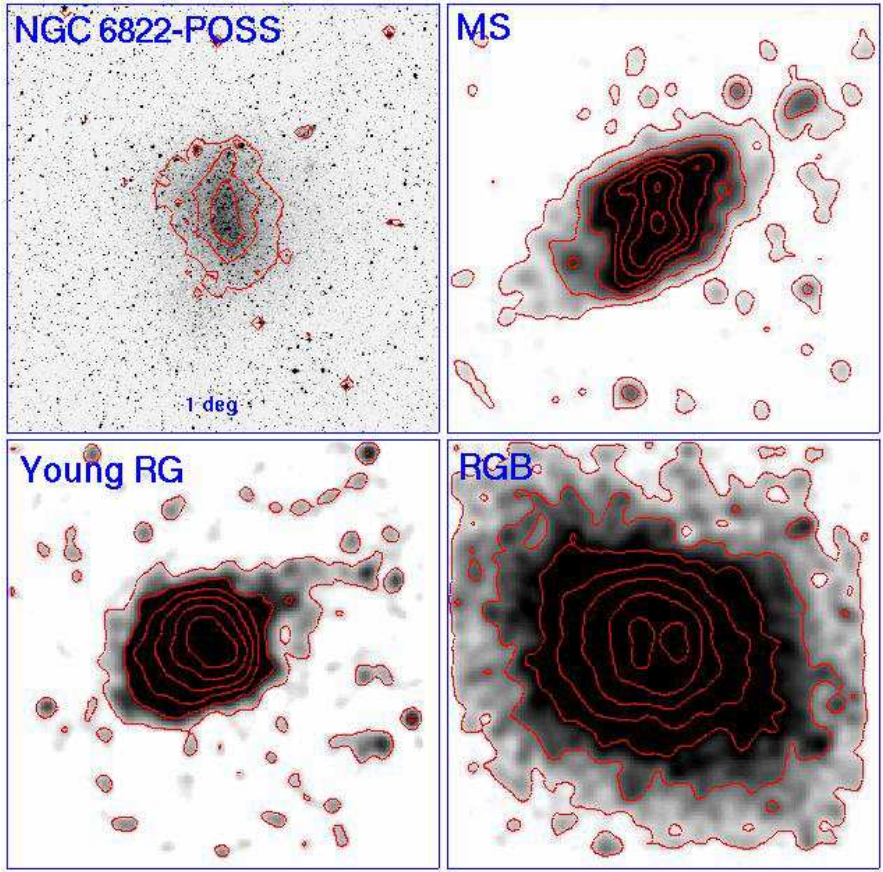}
  \caption{A grayscale map of the digitized POSS  of NGC 6822 (upper left) and the
surface number density maps of the different populations in NGC 6822
(upper right: young main sequence stars;
lower left: young red giants, and lower right: red giant branch stars). Each panel covers
a field of 1 deg $\times$ 1 deg. North is up, and east to the left.
} 
\label{fig2}
\end{figure}

\begin{figure}
\includegraphics[scale=0.50]{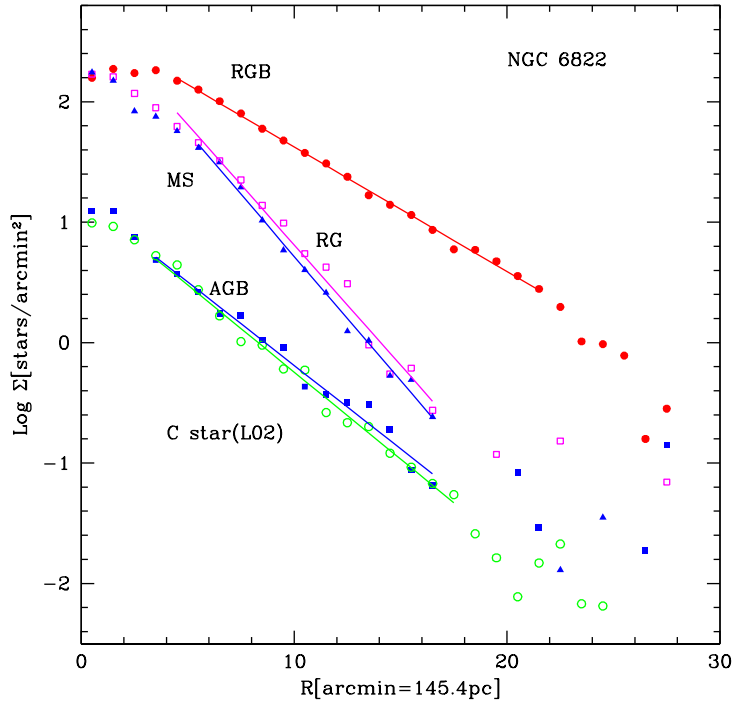}
  \caption{Radial variation of the stellar surface number density of the different populations in NGC 6822
(as marked in Figure~1). Filled circles: old red giants (RGB), open squares: young red giants (RG),
filled triangles: young main sequence stars and blue supergiants (MS), filled squares: asymptotic giant branch (AGB),
 and open circles: carbon stars (Letarte et al. 2002).
Solid lines represent power-law fits to the data.}
\label{fig3}
\end{figure}

\begin{acknowledgments}
This work is supported in part by the ABRL grant (R14-2002-058-010000-0) and the BK21 program.
\end{acknowledgments}

\begin{discussion}

\discuss{Gallagher}{It is known that there is significant differential reddening over the field of NGC 6822. Is there any possibility that the offset between the centers of the old RGB and young RG distribution may be due to differential reddening effect?}

\discuss{Lee}{There is some differential reddening over the field of NGC 6822, but it is not
large enough to affect our results.}

\end{discussion}

\end{document}